\documentclass[aps,prl,twocolumn,groupedaddress,showpacs,showkeys]{revtex4-1}

\usepackage{graphicx}
\usepackage{float}
\usepackage{amsmath}
\usepackage{xfrac}
\usepackage{upgreek}
\usepackage{bm}

\usepackage[version=3]{mhchem}
\DeclareMathOperator\dif{d\!}

\newcommand\ts{\tau_\mathrm{s}}
\newcommand\tsau{\tau_\mathrm{s,Au}}
\newcommand\tspt{\tau_\mathrm{s,Pt}}
\newcommand\tsta{\tau_\mathrm{s,Ta}}
\newcommand\tsal{\tau_\mathrm{s,Al}}
\newcommand\tscu{\tau_\mathrm{s,Cu}}
\newcommand\tsg{\tau_\mathrm{s,Graphene}}
\newcommand\tssi{\tau_\mathrm{s,Si}}
\newcommand\tp{\tau_p}
\newcommand\rc{r_\mathrm{C}}
\newcommand\rn{r_\mathrm{N}}

\newcommand\rsi{r_\mathrm{SI}}
\newcommand\gc{g_\mathrm{C}}
\newcommand\gcup{g_\mathrm{C\uparrow}}
\newcommand\gcdn{g_\mathrm{C\downarrow}}
\newcommand\ls{l_\mathrm{s}}
\newcommand\lsn{l_\mathrm{sN}}

\newcommand\thsh{\theta_\mathrm{SH}}
\newcommand\roun{\rho_\mathrm{N}}
\newcommand\roupt{\rho_\mathrm{Pt}}
\newcommand\routa{\rho_\mathrm{Ta}}

\newcommand\f{v_{1\omega}}
\newcommand\ff{v_{2\omega}}
\newcommand\w{_{1\omega}}
\newcommand\ww{_{2\omega}}

\newcommand\umo{\ \mathrm{m\Omega}}
\newcommand\uuocm{\ \mathrm{\mu \Omega cm}}
\begin{document}


\title{Determination of spin relaxation times in heavy metals via 2nd harmonic spin injection magnetoresistance}

\author{C. Fang}
\affiliation{Institute of Physics, Chinese Academy of Sciences, Beijing, 100190, China.}
\affiliation{University of Chinese Academy of Sciences, Beijing 100049, China.}
\author{C. H. Wan}
\email{wancaihua@iphy.ac.cn}
\affiliation{Institute of Physics, Chinese Academy of Sciences, Beijing, 100190, China.}
\affiliation{University of Chinese Academy of Sciences, Beijing 100049, China.}
\author{A. Hoffmann}
\affiliation{Material Science Division, Argonne National Laboratory, 9700 S. Cass Avenue, Lemont, IL 60439.}
\author{X. M. Liu}
\affiliation{Department of Physics, Shanghai University, Shanghai 200444, China.}
\author{B. S. Yang}
\affiliation{Institute of Physics, Chinese Academy of Sciences, Beijing, 100190, China.}
\affiliation{University of Chinese Academy of Sciences, Beijing 100049, China.}
\author{J. Y. Qin}
\affiliation{Institute of Physics, Chinese Academy of Sciences, Beijing, 100190, China.}
\affiliation{University of Chinese Academy of Sciences, Beijing 100049, China.}
\author{B. S. Tao}
\affiliation{Institute of Physics, Chinese Academy of Sciences, Beijing, 100190, China.}
\affiliation{University of Chinese Academy of Sciences, Beijing 100049, China.}
\author{H. Wu}
\affiliation{Institute of Physics, Chinese Academy of Sciences, Beijing, 100190, China.}
\affiliation{University of Chinese Academy of Sciences, Beijing 100049, China.}
\author{X. Zhang}
\affiliation{Institute of Physics, Chinese Academy of Sciences, Beijing, 100190, China.}
\affiliation{University of Chinese Academy of Sciences, Beijing 100049, China.}
\author{Z. M. Jin}
\affiliation{Department of Physics, Shanghai University, Shanghai 200444, China.}

\author{X. F. Han}
\email{xfhan@iphy.ac.cn}
\affiliation{Institute of Physics, Chinese Academy of Sciences, Beijing, 100190, China.}
\affiliation{University of Chinese Academy of Sciences, Beijing 100049, China.}


\date{\today}

\begin{abstract}
Abstract: In tunnel junctions between ferromagnets and heavy elements with strong spin orbit coupling the magnetoresistance is often dominated by tunneling anisotropic magnetoresistance (TAMR). This makes conventional DC spin injection techniques impractical for determining the spin relaxation time ($\ts$). Here, we show that this obstacle for measurements of $\ts$ can be overcome by 2nd harmonic spin-injection-magnetoresistance (SIMR). In the 2nd harmonic signal the SIMR is comparable in magnitude to TAMR, thus enabling Hanle-induced SIMR as a powerful tool to directly determine $\ts$. Using this approach we determined the spin relaxation time of Pt and Ta and their temperature dependences. The spin relaxation in Pt seems to be governed by Elliott-Yafet mechanism due to a constant resistivity$\times$spin relaxation time product over a wide temperature range.
\end{abstract}
\maketitle

The large applied potential of spin-orbit-torques for magnetic random access memory has stimulated intensive interest in investigating spin orbit coupling (SOC) in heavy metals such as Pt and Ta~\cite{ref01,ref02,ref03,ref04,ref05,ref06,ref07,ref08,ref09,ref10,ref11}. Their spin Hall angle ($\thsh$), spin diffusion length ($\ls$) and spin relaxation time ($\ts$), which influence switching efficiency are important parameters for determining their effectiveness, but especially the latter two are experimentally hard to assess. Accurate determination of $\ts$ could also help to identify the spin relaxation mechanisms~\cite{ref12}. Though $\thsh$ and $\ls$ have been measured by spin pumping~\cite{ref13,ref14,ref15,ref16,ref17} and 2nd harmonic Hall measurement~\cite{ref18,ref19,ref20}, $\ts$ of Pt and Ta is rarely reported. In principle, $\ts=\ls^2/D$, with $D$ being the diffusion constant, which is also difficult to determine independently.

Electron spin resonance (ESR) has been a standard technique to measure the spin relaxation time of bulk light metals~\cite{ref21}. However, it is not suitable for ultrathin films~\cite{ref22,ref23}. In addition, Elezzabi et al.~\cite{ref24} developed a time-resolved optical technique to directly measure the spin relaxation process in Au to be $\tsau=(45\pm5)$ ps. However, this method is not suitable for heavy metals such as Pt, Ta and W with short $\ts$ \cite{ref25}. Recently, Dyakonov\cite{ref26} theoretically, then V\'{e}lez et al.~\cite{ref27} and Wu et al.~\cite{ref28} experimentally demonstrated a so-called Hanle magnetoresistance (MR) effect in Pt and Ta: a spin accumulation at the sample boundaries caused by the spin Hall effect is dephased by a magnetic field via the Hanle effect, which results in an additional positive MR. This electrical method can be applied to estimate $\ts$ from the magnetic field dependence\cite{ref27,ref28}. Using this approach $\tspt=$1.9 ps was determined for Pt/\ce{SiO2} and 0.61 ps for Pt/YIG~\cite{ref28}.

In fact, spin injection experiments in nonlocal spin valves~\cite{ref29,ref30,ref31,ref32,ref33,ref34,ref35} and 3-terminal geometries~\cite{ref36,ref37,ref38,ref39,ref40} are both powerful tools in measuring $\ts$ in metals and semiconductors. In these experiments, ferromagnetic layer (FM)/tunnel barrier/nonmagnetic layer (NM) junctions are adopted to both inject a non-equilibrium spin accumulation and simultaneously determine their magnitude.  These measurement were used to determine spin relaxation times in a wide variety of materials, e.g., $\tssi$=55 - 285 ps for heavily doped silicon~\cite{ref40}; $\tsg >$ 1 ns for graphene/BN~\cite{ref41}; $\tsal$=110 ps for aluminum~\cite{ref29}, $\tscu$=22 ps for copper~\cite{ref42} and $\tsau$=45 ps for gold~\cite{ref32}.

However, it is impractical to apply these spin injection experiments to measure $\ts$ in heavy metals with strong SOC for at least two reasons. First, $\ls$ in this case is so short (about several nanometers) that the preparation of nonlocal spin valves with comparable dimensions is beyond current lithography capabilities. Second, the real contact resistance is $r=\rc + \rsi$, where $\rsi$ and $\rc$ are the contact resistance induced by spin injection (SI) and the original contact resistance without $\rsi$, respectively. Here $\rsi$ equals to $\rn\rc/(\rn+\rc)$ and the spin resistance in the NM layer $\rn$ is defined as $\roun\lsn$. $\roun$ and $\lsn$ are the resistivity and spin relaxation length of NM, respectively. For this discussion we ignore the influence of spin resistance in FM on $\rsi$ due to the small values of $\ls$ in FM. Because $\rn\ll\rc$ for metals, $r\approx\rc+\rn$. As one increases a field perpendicular to the spin polarization in the NM, the spin accumulation dephases, resulting in a vanishing $\rn$ due to the Hanle effect. This gives rise to a $MR\equiv[r(H\rightarrow\infty)-r(0)]/r(0)=-\rn/(\rc+\rn)\approx-\rn/\rc<0$. This negative spin-injection-induced MR (SIMR) has been utilized in 3-terminal geometries to measure $\ts$ in semiconductors~\cite{ref36,ref37,ref38,ref39} but is negligible in metallic systems, since $\rn\ll\rc$ by several orders of magnitude. Besides, $\rc$ can also exhibit a field dependence due to SOC in FM/Barrier/NM junctions~\cite{ref43,ref43.1}. This so-called tunneling anisotropy MR (TAMR)~\cite{ref44} further complicates the analysis.

In this Letter, we will show that even with a 3-terminal geometry, SIMR can be clearly observed by 2nd harmonic voltage measurements, since TAMR only dominates the 1st harmonic voltages. We adopted this method to determine $\ts$ in Pt and Ta and also their corresponding temperature dependences.

First we discuss the basic concept of these measurements. The tunneling conductance $\gc=1/\rc$ is composed by counterparts for opposite spin channels, $\gc=\gcup+\gcdn$. Here we have already neglected $\rn$ in the contact resistance due to the fact that $\rn\ll\rc$. Spin injection into the NM or spin extraction from NM induces a non-equilibrium spin accumulation $\mu_\mathrm{N}$ in NM, which increases or decreases Fermi levels of opposite spin channels. This can further lead to a change of $\gc$ by $\triangle\gc=\frac{\dif\gcup}{\dif E}\mu_\mathrm{N}-\frac{\dif\gcdn}{\dif E}\mu_\mathrm{N}=\frac{\dif(\gcup-\gcdn)}{\dif E}\mu_\mathrm{N}$. The spin accumulation is given by $\mu_\mathrm{N}=p\rn j$, where $p$ and $j$ are the tunneling spin polarization and current density across the junction~\cite{ref45}. Thus $\triangle\gc=\alpha p\rn j$ with $\alpha\equiv\frac{\dif(\gcup-\gcdn)}{\dif E}$. The voltage across the junction $v=\rc j$ is then
\begin{equation}
v=\frac{1}{(g_\mathrm{C,0}+\triangle\gc)}j\approx ( \frac{1}{g_\mathrm{C,0}}-\frac{\triangle\gc}{g_\mathrm{C,0}^2}) j=\frac{1}{g_\mathrm{C,0}}j-\frac{\alpha p\rn }{g_\mathrm{C,0}^2}j^2
\end{equation}

Here $g_\mathrm{C,0}$ is the contact conductance at zero current, or $v=r_\mathrm{C,0}j-\alpha p\rn r_\mathrm{C,0}^2 j^2$ with $r_\mathrm{C,0}$ being the contact resistance at zero current. Note that $r_\mathrm{C,0}$ does not contain SIMR. Assuming that $r_\mathrm{C,0}=r_\mathrm{C,00}(1+\mathrm{TAMR})$ and $\rn=r_\mathrm{N,0}(1+\mathrm{SIMR})$, results in $v=r_\mathrm{C,00}(1+\mathrm{TAMR})j-\alpha p r_\mathrm{N,0} r_\mathrm{C,00}^2\mathrm{(1+SIMR)(1+TAMR)}^2 j^2$, where $r_\mathrm{C,00}$ and $r_\mathrm{N,0}$ are the contact resistance and spin resistance at $H=0$ and $j=0$, respectively. This equation can be further reduced considering TAMR$\ll$1 and SIMR$\ll$1:

\begin{widetext}
\begin{equation}
v\approx r_\mathrm{C,00}(1+\mathrm{TAMR})j-\alpha p r_\mathrm{N,0} r_\mathrm{C,00}^2(1+\mathrm{SIMR+2TAMR})j^2
\end{equation}
\end{widetext}

In practice, an AC current $j=j_0\sin(\omega t)$ satisfying $\triangle\gc<g_\mathrm{C,0}/10$ was selected to make the above Taylor expansion reasonable. Thus $\f= r_\mathrm{C,00}(1+\mathrm{TAMR})j_0$ has no explicit dependence on SIMR while $\ff=\frac{1}{2} \alpha p r_\mathrm{N,0} r_\mathrm{C,00}^2(1+\mathrm{SIMR+2TAMR})j_0^2$ has a dependence on both SIMR and 2TAMR. They also differ in phase by 90$^\circ$. We would expect that TAMR dominates in $\f$ while SIMR becomes comparable to the TAMR and thus observable in $\ff$ as shown in the following experiments.

\begin{figure}[thb!]
\includegraphics[width=9cm]{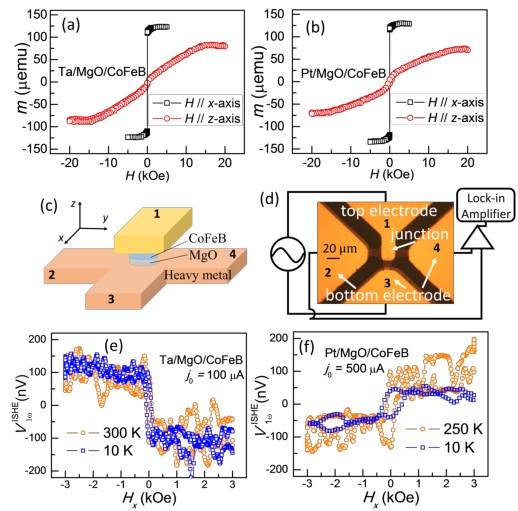}%
\caption{\label{fig1}(Color online) (a) and (b) Magnetic moment $m$ vs $H$ curve of Ta/ MgO/CoFeB and Pt/ MgO/CoFeB film. (c) Schematic of heavy metal/ MgO/\ce{Co40Fe40B20} junctions. Top electrode 1 and bottom electrodes 2, 3 and 4 are on opposite sides of 40-nm MgAlO$_x$ around the tunnel junction area. (d) The ISHE measurement setup applying an AC current between E1 and E3 and detecting the voltage between E2 and E4 with preamplifier and lock-in amplifier. (e) and (f) 1st harmonic ISHE voltage of Ta/ MgO/CoFeB and Pt/ MgO/CoFeB. High temperature (orange circle) or low temperature (blue) data are shown together for the Ta and Pt stacks, respectively. The current amplitude is 100 $\mathrm{\mu A}$ for Ta and 500 $\mathrm{\mu A}$ for Pt. Opposite field dependencies (e) and (f) indicate different signs of $\thsh$ of Ta and Pt.}
\end{figure}

Stacks of \ce{SiO2}//Ta(10) or Pt(10)/MgO(2)/\ce{Co40Fe40B20}(4)/Ta(5)/Ru(7) (thickness in nm) provided by Singulus Technologies AG were deposited via magnetron-sputtering and then post-annealed with a magnetic field of 1 $\mathrm{T}$ along the $x$-axis at 300 $^o$C for 1 hour to induce an easy axis along the $x$-axis. $M$-$H$ curves acquired by vibrating sample magnetometer (Microsense) showed in-plane magnetic anisotropy for both Ta/MgO/CoFeB and Pt/MgO/CoFeB stacks [Fig. 1(a) and (b)]. The anisotropy field of each sample is about 15 kOe along the $z$-axis, while the easy axis is along the $x$-axis. $H_x$ smaller than 1 kOe is sufficient to align the magnetization along the easy axis.

The extended films were then processed into magnetic tunneling junctions by ultraviolet lithography and argon ion etching. The junctions had one top electrode (E1) and three bottom ones (E2, E3 and E4) [Fig. 1(c) and (d)]. The size of the junctions was 6 $\mathrm{\mu m}\times$6 $\mathrm{\mu m}$. Ta/MgO/CoFeB or Pt/MgO/CoFeB junctions were surrounded by MgAlO$_x$ for protection and also for isolating E1 from the remaining electrodes. Magnetotransport properties were measured in a physical property measurement system (Quantum Design-9T). To measure the inverse spin Hall effect (ISHE) of the bottom electrodes, an AC current with sine wave and $f=\omega/\pi=8.7$ Hz was applied between E1 and E3 using a Keithley 6221 and the 1st harmonic voltage $V\w$ between E2 and E4 was firstly pre-amplified (Stanford Research, SR560) and then picked up by a lock-in amplifier (SR830) [Fig. 1(d)].

In this setup, spin-polarized current was perpendicularly injected from the FM to the NM layer. Their spin orientation was along the $x$-axis at $|H_x|>$500 Oe. Then a voltage in the open circuit can be detected along the $y$-axis due to the ISHE. The field dependences of the 1st harmonic voltage $V\w^\mathrm{ISHE}$ between E2 and E4 in Ta and Pt junctions are illustrated in Fig. 1(e) and (f). The sign of $V\w^\mathrm{ISHE}$ reverses as expected with reversed sign of $H_x$. $V\w^\mathrm{ISHE}$ has opposite signs in the Ta and Pt due to their opposite $\thsh$~\cite{ref47,ref48}, which indicates successful spin injection into the bottom heavy metal layer. Similar ISHE behaviors in both junctions have also been observed near room temperature. The maximum $V\w^\mathrm{ISHE}/j_0$ of Ta and Pt junctions is about 1 $\umo$ and 0.1 $\umo$ at 300 K, which is in the same order of magnitude as in Ref. ~\cite{ref46}.

\begin{figure}[thb!]
\includegraphics[width=9cm]{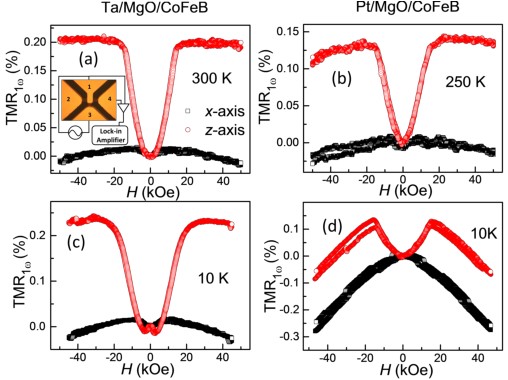}%
\caption{\label{fig2}(Color online) TMR obtained from 1st harmonic voltage with the 3-terminal (3T) measurement setup applying AC currents between E1 and E3 and detecting the voltages between E1 and E4 in the inset at high temperature (a) 300 K for Ta/MgO/CoFeB, (b) 250 K for Pt/MgO/CoFeB or low temperature 10 K for (c) Ta/MgO/CoFeB or (d) Pt/MgO/CoFeB. The external field is either in plane along $x$-axis (black square) or out of plane along $z$-axis (red circle). The currents are identical as in Fig. 1. (e) and (f), 100 $\mathrm{\mu A}$ for Ta/MgO/CoFeB [(a) or (c)] and 500 $\mathrm{\mu A}$ for Pt/MgO/CoFeB [(b) or (d)].}
\end{figure}

3-terminal MR measurements are further performed on both Ta and Pt junctions. We have first detected the 1st harmonic voltage $V\w^\mathrm{3T}$ between E1 and E4 with an AC current applied between E1 and E3 [inset of Fig. 2(a)]. TMR$\w$ is defined as [$V\w^\mathrm{3T}(H)$-$V\w^\mathrm{3T}(0)$]/$V\w^\mathrm{3T}(0)$ and its field dependences is shown in Fig. 2(a)-(d). The MR originates from the tunneling junction instead of the anisotropy magnetoresistance (AMR) of the CoFeB layer. Direct measurements of AMR of the Ta and Pt showed negligible field dependence in the 1st harmonic measurements. AMR only appears in the DC measurement, whose value is negligibly small, only less than 0.05$\%$ at 10 K\cite{ref48.1}. Except Thus, the TMR is mainly attributed to anisotropic tunneling magnetoresistance (TAMR) of the CoFeB/MgO/heavy metal junctions, and we use TAMR instead of TMR in the following analysis.

At high temperature, TAMR$\w^z$ first quadratically increases as $H_z$ increases from zero in both Ta and Pt junctions [Fig.2(a) and (b)] and later gradually saturates at 0.20$\%$ for Ta and 0.14$\%$ for Pt junction as $H_z$ approaches 15 kOe which is also the anisotropy field of the CoFeB layer. Further increasing $H_z$ leads to a MR reduction for both junctions. When $H_x$ is applied, TAMR$\w^x$ increases only by about 0.01$\%$ and then decreases gradually toward the negative MR. Note that TAMR$\w^z$ is much larger than TAMR$\w^x$. $H_z$ aligns the magnetization from in-plane to out-of-plane, which subsequently changes the density of state of the interfacial FM layer via SOC and results in TAMR as predicted theoretically~\cite{ref44,ref49}. The phenomenon TAMR$\w^z>$TAMR$\w^x$ is consistent with Ref.\cite{ref50}, since $H_x$ keeps the magnetization along the easy axis, and consequently TAMR$\w^x$ varies little.

Similar behaviors are also observed at 10 K, except for larger saturation fields and slightly larger TAMR$\w^z$  values [Fig. 2(c) and (d)]. The negative MR ,which depends on applied field instead of magnetization, is also observed at 10 K. Its origin is still unknown and beyond the scope of this study. The only remarkable difference between 10 K and high temperature is that a small negative MR (about -0.014$\%$) appears at low $H_z$ in the Ta junction [Fig.2(c)]. This negative MR exhibits a similar field dependence as the Hanle-effect-induced SIMR discussed below. Thus we attribute it to spin injection into Ta. This SIMR$\w$ should have been negligibly small due to the fact $\rn\ll\rc$. In fact, it turns out to be unobservable in the Pt junction or at high temperatures.
It might be possible that inhomogeneities of the MgO layer result in a significant reduction of the effective tunneling area and smaller $\rc$ in the Ta junction. This may lead to a reemerging of SIMR$\w$ although SIMR$\w$ is still one order smaller than TAMR$\w^z$. Inhomogeneous current distribution due to the resistance of the nonmagnetic layer within the junction area could reduce the measured tunneling resistance below the real tunneling resistance by about 10.8$\%$ and 4.5$\%$ for Ta and Pt junctions respectively due to device geometry as well as inhomogeneous current distribution within the junction~\cite{ref50.1,ref50.2}. However, this would not affect the injected spins and their dephasing process in the heavy metal layers. Therefore, this resistance adjustment would not physically influence the field dependence of the TAMR and the SIMR effects which is the basis of estimating the spin relaxation times.

$V\ww^\mathrm{3T}$ was detected in the same setup as shown in the inset of Fig. 2(a). The only difference is that the 2nd harmonic voltage with 90$^\circ$ phase shift was measured with the lock-in amplifier. As shown in Eq. (2), SIMR should be comparable to TAMR within a factor of 2 for the 2nd harmonic signal. Thus this method renders Hanle and inverted Hanle effect signals induced by SIMR detectable even in the presence of a TAMR background (Fig.3).

\begin{figure}[thb!]
\includegraphics[width=9cm]{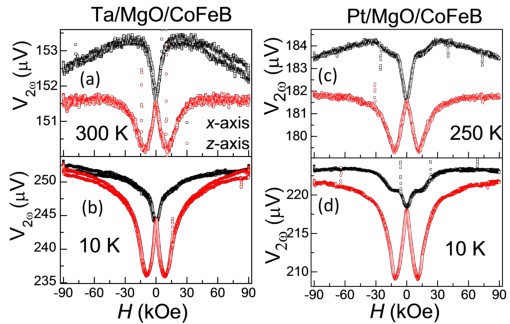}%
\caption{\label{fig3}(Color online) 2nd harmonic voltage with the 3-terminal (3T) measurement setup for Ta/MgO/CoFeB at (a) 300 K or (b) 10 K, and for Pt/MgO/CoFeB at (c) 250 K or (d) 10 K. The magnetic field was applied along the $x$-axis (black square) for inverted Hanle measurement or the $z$-axis (red circle) for Hanle measurement.}
\end{figure}

The field dependence of $V\ww^\mathrm{3T}$ at 300 K or 250 K for Ta and Pt junction is shown in Fig. 3(a) and (b). For small $H_z$, the magnetization is still aligned along the easy axis. An AC current injects (extracts) spins into (from) NM and leads to a non-equilibrium spin accumulation, which conversely influences tunneling resistance and contributes an additional $V\ww$. A vertical $H_z$ can dephase the spin accumulation via the Hanle effect and therefore diminish the additional $V\ww$, leading to a negative MR with a Lorentzian shape in the 2nd harmonic signal. This Hanle dephasing is the same as established by Silsbee~\cite{ref51} for DC measurement. It is worth noting that TAMR$\ww$ and SIMR$\ww$ contribute to a positive and negative MR, respectively. In addition, TAMR$\ww$ has a $H_z^2$ dependence at low field according to our results in Fig. 2, while SIMR$\ww$ exhibits a Lorentzian-shape dependence. By fitting $V\ww$ vs. $H_z$ curves with a Lorentzian function plus a $H_z^2$ function, we can obtain a spin relaxation time $\ts=e/(mB_0)$ with the electron charge $e$, electron mass $m$ and $B_0$ being the half width at half maximum of the Lorentzian fitting. $\ts$ is (7.8$\pm$1.6) ps at 300 K and (13.1$\pm$0.6) ps at 10 K for Ta [Fig.3(a) and (b)]. By further increasing $H_z$ beyond 10 kOe, $V\ww^\mathrm{3T}$ increases due to both tilting of magnetization and the concomitant TAMR contribution.

In contrast, $H_x$ avoids dephasing of the spin polarization along $x$, and therefore extends spin relaxation process and finally causes a positive MR in small field. This picture accounts for the inverted Hanle effect~\cite{ref40}. A similar positive SIMR also occurs for the 2nd harmonic signal (Fig.3). Besides, $V\ww$ exhibits a $H_{z/x}$ dependence at high fields, especially at 10 K, but the origin of this field dependence is unclear at this point. The Hanle signal in Fig.3 (c) and (d) results in $\ts$ (5.0$\pm$1.5) ps at 250 K and (7.3$\pm$0.6) ps at 10 K for Pt. The inverted Hanle SIMR shows similar behavior for Ta. More than 4 devices are measured to estimate the $\ts$ for each type of stacks. The data for the other devices are attached in the Supplementary Information\cite{ref48.1}.

\begin{figure}[thb!]
\includegraphics[width=9cm]{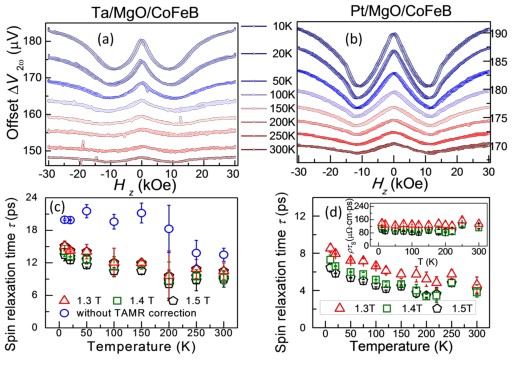}%
\caption{\label{fig4}(Color online) Temperature dependence of the 2nd harmonic voltage of Hanle measurements for (a) Ta/MgO/CoFeB and (b) Pt/MgO/CoFeB from 10 K to 300 K. And temperature dependence of spin relaxation time (c) for Ta/MgO/CoFeB and (d) for Pt/MgO/CoFeB acquired via fitting the data with a Lorentzian curve plus a parabolic function for the TAMR correction applied in different field ranges $\pm$13 Oe (red triangle), $\pm$14 Oe (olive square) and $\pm$15 Oe (black pentagon) or without the parabolic function fitting (blue circle). Inset in (d) shows that $\ts\rho$ remains nearly constant from 300 K to 10 K for all fitting ranges.}
\end{figure}

In order to investigate the temperature ($T$) dependence of $\ts$, we have conducted the 2nd harmonic SIMR measurement in a Hanle geometry at different temperatures [Fig.4(a) and (b)]. As $T$ decreases from 300 K to 10 K, the Hanle-effect-induced $\triangle V\ww$ grows significantly by nearly one order of magnitude. In order to examine whether the field range for selecting the data affects $B_0$, we have tried different ranges ($\pm$13 kOe, $\pm$14 kOe and $\pm$15 kOe) for the fitting. The $T$ dependence is basically the same for different fitting ranges. Their variance is less than 2 ps for both materials. Taking the $\pm$14 kOe fitting range, $\ts$ in Ta gradually decays from (13.1$\pm$0.6) ps at 10 K to (7.8$\pm$1.6) ps at 300 K. In contrast, if the TAMR correction is ignored in the fitting $\ts$ stays at 20 ps below 150 K and then decays to 14 ps at 300 K. These values are not only 50$\%$ higher than those with TAMR correction but also exhibits an unreasonable $T$ dependence. Thus the TAMR correction is indispensable. $\ts$ of Pt and Ta is about 10 ps or below. These values are 1-3 orders smaller than $\ts$ in light metals or semiconductors, consistent with the trend that elements with larger atomic number have stronger SOC. $\tspt$ is about half of $\tsta$ at all temperatures in our experiment and much smaller than $\tsau$ of 45 ps. Here $\tspt$=(5.0$\pm$1.5) ps at 250 K is about twice of 1.9 ps measured by Hanle MR, which might be caused by lower resistivity in the former Pt and different film thickness in the two experiments.  In our experiment, $\roupt$=24.4$\uuocm$ at 300 K, while it is 58$\uuocm$ in Ref.~\cite{ref28}. $\ts \rho$ appears to be a constant for these two samples. The $T$ dependence of $\roupt$ is also measured. For resistivity measurement, the top structure MgO/CFB/capping layers in the Pt/MgO/CFB stacks is etched away. $\roupt$ decreases weakly with decreasing temperature and $\ts \rho$ in Pt is nearly a constant from 300 K to 10 K for all the fitting ranges [inset in Fig.4(d)]. The momentum relaxation time $\tp$ is inversely proportional to $\rho$. Thus $\ts/\tp$ is also a constant, which indicates that the spin relaxation in Pt is governed by Elliott-Yafet mechanism ~\cite{ref12}. We also applied a THz technique~\cite{ref52} to directly measure momentum relaxation time and resistivity of Pt with 30 nm thickness, which gives $\tp$=(5$\pm$3) fs and $\roupt$=16$\uuocm$ at 300 K. Assuming that $\tp$ is proportional to 1/$\roupt$, $\tp$ in Pt/MgO/CFB is thus around 2.7 fs. Therefore the spin flip probability of each scattering $\tp/\ts$ is around 7$\times$10$^{-4}$   for Pt at 300 K.

Our $\routa$ is about 342$\uuocm$ at 300 K, much larger than those reported for the resistivity of $\alpha$-phase and even $\beta$-phase Ta~\cite{ref53,ref54}, which might be due to oxidation of Ta after the top structure is etched. Therefore $\routa$ vs. $T$ is not used here for examining the spin relaxation mechanism.

In conclusion, TAMR$\w$ dominates the 1st harmonic 3-terminal MR measurement while SIMR$\ww$ becomes significant compared to the TAMR$\ww$ background and turns out to be much easier measured in the 2nd than in the 1st harmonic signal. This renders conventional 3-terminal FM/barrier/NM devices suitable for directly measuring the spin relaxation time $\ts$ of heavy metals without complications from proximity effects~\cite{ref55,ref56,ref57,ref58} that occur, when the heavy metal is in direct contact with a ferromagnet. ISHE is also observed, which proves successful spin injection into Ta and Pt. By fitting Hanle curves with a Lorentzian function plus a parabolic TAMR background, we have obtained $\ts$ of Ta and Pt. The $\ts$ for both materials exhibits a small increase from 300 K to 10 K, such that $\ts$ is about (7.8$\pm$1.6) ps and (5.0$\pm$1.5) ps for Ta and Pt at high temperature while it is about (13.1$\pm$0.6) ps and (7.3$\pm$0.6) ps at 10 K, respectively. Since $\ts \rho$ stays constant at all temperatures, the spin relaxation in Pt seems to be dominated by the Elliott-Yafet mechanism. This experimental approach provides an electrical manner to directly quantify spin relaxation time of heavy metals, which have been elusive from conventional SIMR or optical measurements. Furthermore, there is no physical limitation for this method to be generalized to other light metals and semiconductors.


\section{acknowledgments}
\begin{acknowledgments}
This work was supported by the 863 Plan Project of Ministry of Science and Technology (MOST) (Grant No. 2014AA032904), the MOST National Key Scientific Instrument and Equipment Development Projects [Grant No.2011YQ120053], the National Natural Science Foundation of China (NSFC) [Grant No. 11434014, 51229101, 11404382] and the Strategic Priority Research Program (B) of the Chinese Academy of Sciences (CAS) [Grant No.XDB07030200]. The work from Axel Hoffmann contributing to experiment conception and data analysis was also supported by the U.S. Department of Energy, Office of Science, Basic Energy Sciences, Materials Science and Engineering Division. X. M. Liu and Z. M. Jin have contributed to THz measurement. The annealed raw films were provided by Singulus Technologies AG.
\end{acknowledgments}

%

\end{document}